\documentclass[twoside]{article}
\usepackage[utf8]{inputenc}
\usepackage[mathletters]{ucs}
\usepackage{amsmath,amssymb}
\usepackage{qic,epsfig}

\usepackage{graphicx}
\newcommand{\<}{\langle}
\renewcommand{\>}{\rangle}
\newcommand{\I}{{\rm I}}

\usepackage{marginnote}

\def\id{{\operatorname{id}}}
\def\I{{\rm I}}
\newcommand{\tr}{{\rm Tr}}

\def\bei{\begin{itemize}}
\def\eei{\end{itemize}}
\def\be{\begin{equation}}
\def\ee{\end{equation}}

\def\Mt{\tilde M}
\newtheorem{fact}{Fact}
\newtheorem{corrolary}{Corrolary}
\newtheorem{proposition}{Proposition}

\newenvironment{examples}%
{\par\addvspace{\medskipamount}\noindent\textbf{Examples.}\hspace{1ex}}%
{\par\medskip}

\newenvironment{remark}%
{\par\addvspace{\medskipamount}\noindent\textbf{Remark.}\hspace{1ex}}%
{\par\medskip}

\DeclareUnicodeCharacter{981}{\phi}
\DeclareUnicodeCharacter{348}{{\hat{S}}}
\DeclareUnicodeCharacter{8224}{{\dagger}}
\newcommand\kEXT{\ensuremath{\textrm{EXT}_k}}
\newcommand\M{M_{ab}^α}

\usepackage{color}

\begin{document}
\setlength{\textheight}{8.0truein}    %FOR 2ND PAGE ONWARDS

\runninghead{Entanglement distillation by extendible maps}
            {\L{}. Pankowski, F.\ G.\ S.\ L.\ Brand\~ao, 
             M.\ Horodecki, G. Smith }

\normalsize\textlineskip
\thispagestyle{empty}
\setcounter{page}{1}

%\copyrightheading{Vol.}{No.}{Year}{Page Nos.}
\copyrightheading{0}{0}{2003}{000--000}

\vspace*{0.88truein}

\alphfootnote

\fpage{1}

\centerline{\bf
%%%%%%%%%%%%%%%%%%%%%
%Put in titiles here
%%%%%%%%%%%%%%%%%%%%%
  ENTANGLEMENT DISTILLATION BY EXTENDIBLE MAPS}
\vspace*{0.37truein}
\centerline{\footnotesize
%%%%%%%%%%%%%%%%%%%%%%%%%%%%%%%%%%%%
%put authors' name and address here
%%%%%%%%%%%%%%%%%%%%%%%%%%%%%%%%%%%%
\L{}UKASZ PANKOWSKI}
\vspace*{0.015truein}
\centerline{\footnotesize\it
  Institute of Theoretical Physics and Astrophysics,
  University of Gda\'nsk, ul. Ba\.zy\'nskiego 1a}
\baselineskip=10pt
\centerline{\footnotesize\it
  Gda\'nsk, 80-952, Poland}

\vspace*{10pt}
\centerline{\footnotesize
FERNANDO G.\ S.\ L.\ BRAND\~AO}
\vspace*{0.015truein}
\centerline{\it
Departamento de Fisica, Universidade Federal de Minas Gerais}
\baselineskip=10pt
\centerline{\footnotesize\it
Belo Horizonte, 30123-970, Brazil}

\vspace*{10pt}
\centerline{\footnotesize
  MICHA\L{} HORODECKI}
\vspace*{0.015truein}
\centerline{\footnotesize\it
  Institute of Theoretical Physics and Astrophysics,
  University of Gda\'nsk, ul. Ba\.zy\'nskiego 1a}
\baselineskip=10pt
\centerline{\footnotesize\it
  Gda\'nsk, 80-952, Poland }

\vspace*{10pt}
\centerline{\footnotesize
 GRAEME SMITH}
\vspace*{0.015truein}
\centerline{\footnotesize\it
IBM T.J. Watson Research Center, Yorktown Heights}, 
\baselineskip=10pt
\centerline{\footnotesize\it
  NY 10598, USA}

\vspace*{0.225truein}
\publisher{(received date)}{(revised date)}

\vspace*{0.21truein}

%% \abstracts{first paragraph}{second paragraph}{third paragraph}
%% If there is only one paragraph, just keep the second and third empty 
%% like the following one 
\abstracts{
%%%%%%%%%%%%%%%%%%%%
% put abstract here
%%%%%%%%%%%%%%%%%%%%
It is known that from entangled states that have positive partial
transpose it is not possible to distill maximally entangled states by
local operations and classical communication (LOCC). A long-standing
open question is whether maximally entangled states can be distilled from
every state with a non-positive partial transpose. In this paper we
study a possible approach to the question consisting of enlarging the class 
of operations allowed. Namely, instead of LOCC operations we consider 
$k$-extendible operations, defined as maps whose Choi-Jamio\l{}kowski 
state is $k$-extendible. We find that this class is unexpectedly powerful - 
e.g. it is capable of distilling EPR pairs even from completely product states. 
We also perform numerical studies of distillation of Werner states by those maps, which show
that if we raise the extension index $k$ simultaneously with the number of
copies of the state, then the class of $k$-extendible operations are
not that powerful anymore and provide a better approximation to the
set of LOCC operations.
}{}{}

\vspace*{10pt}

\vspace*{3pt}
\communicate{to be filled by the Editorial}

\vspace*{1pt}\textlineskip    %) USE THIS MEASUREMENT WHEN THERE IS
   %) A SECTION HEADING
%\vspace*{-0.5pt}
%\noindent
%%%%%%%%%%%%%%%%%%%%%%%%%%%%%%%%
%put the text of the paper here
%%%%%%%%%%%%%%%%%%%%%%%%%%%%%%%%

\section{Introduction}

Entanglement is a quantum resource without analogue in the classical
world. For instance entanglement allows one party (Alice) to teleport an
unknown state of her quantum system to a distant quantum system held
by another party (Bob). Teleportation requires, for every teleported
qubit, that Alice and Bob share an EPR pair and that Alice sends two
bits of information (the result of her measurement) to Bob.

Maximally entangled states are especially valuable as they allow for
perfect teleportation. In practice, however, we can only obtain imperfect
entangled states due to the interaction with the environment. In this
context entanglement distillation methods have been developed
\cite{BBPSSW1996,BDSW1996} obtaining from a large number of
copies of imperfect entangled states a smaller number of maximally
entangled states.

Later \emph{bound entangled} states were discovered \cite{bound}
which, although entangled, cannot be distilled into maximally
entangled states. Entangled states can be divided into so-called
positive partial transpose (PPT) states and non-positive partial
transpose (NPT) ones. It is known that all PPT entangled states are
bound entangled but it is still a long-standing open question whether some of the
NPT states are also bound entangled.

Since this question was raised in \cite{bound}, many attempts to solve
the problem have been made, and several partial results have been
obtained (see
e.g. \cite{DurCLB1999-npt-bound,DiVincenzoSSTT1999-nptbound,Clarisse2006-bound,PankowskiPHH-npt2007,
  BrandaoE10} and \cite{RMPK-quant-ent} for further references). In
particular, it is known that if there are NPT bound entangled states
there must exist NPT bound entangled Werner states
\cite{reduction}. This allows one to focus on the distillability of
Werner states.

The problem of distillation of a given state ρ is the question whether
Alice and Bob having many copies of the state ρ can, using LOCC
operations (Local Operations and Classical Communication), obtain a
smaller number of copies of maximally entangled states.

One of the possible research directions is to allow Alice and Bob to
use a broader class of operations than LOCC. In \cite{WernerPPT} the
class of so-called PPT operations \cite{Rains2001} was considered, and
it was shown, that any NPT state can be distilled by means of such
operations and thus this class is not useful to the question about the
existence of NPT bound entangled states. The same is true for the
class of non-entangling operations \cite{BP08}, as they can distill EPR pairs
from every entangled state. 

In this paper we consider another class of operations, namely the
class of \emph{$k$-extendible operations}. They are an interesting
class of operations since in the limit $k\to\infty$ they converge to
the class of separable operations (and a state is distillable by LOCC
if and only if it is distillable by separable operations). In more
detail $k$-extendible operations are maps whose corresponding
Choi-Jamio\l{}kowski state has a $k$-symmetric extension. The
class of $k$-extendible maps are directly related to the criteria of
entanglement based on $k$-extendability of states, an important method
in the detection of entanglement (see
e.g. \cite{DohertyPS04,Werner1989-sym-ext,Yang2006-sym-ext,NavascuesOP2009-sep,BrandaoCY11a,BrandaoCY11b}).

%We shall not require the operations to be trace-preserving. Our main
%quantity of interest will be fidelity of output with maximally
%entangled state, that can be obtained with some nonzero probability.

Out first result is that $k$-extendible maps are extremely powerful
regarding distillation. Namely, we prove that for any fixed $k$, the
class of $k$-extendible maps can distill any state but the maximally
mixed one, if there are sufficiently many copies available. Second,
even with a single copy of the state, the maps can provide fidelity
$1$ (with nonzero probability) for any state which has a $k$-extendible
state in its kernel. This means, in particular, that the fidelity
achievable by means of such maps is not stable under local embedding
into a larger Hilbert space.

We then analyze the case of Werner states. By use of irreducible
representations of the symmetric group, we obtain analytically the
maximal fidelity achievable for a single copy of Werner states and for
$1$-extendible maps.  The curve of attainable fidelity turns out to be
symmetric with respect to the maximally mixed state, on the interval
joining the symmetric state (which is separable) and the antisymmetric
state (the most entangled Werner state). Thus the action of
$1$-extendible maps is independent to the entanglement of the
state. We then consider a subclass of $k$-extendible maps, which we
call `\emph{measure-and-prepare}' maps (they are all entanglement
breaking channels). For single copy of Werner state and $k=1$ we show
that this subclass gives the same fidelity as the maximization over
all $1$-extendible maps.  We then provide some numerical analysis: We
obtain that for small number $k \leq 4$ of extensions, the symmetry
with respect to maximally mixed state is still present, and it breaks
at $k=5$.

The paper is organized as follows:
In Section \ref{sec:fid} we derive the formula for the achievable fidelity 
by $k$-extendible maps. We also consider a subclass of "measure-and-prepare" maps, 
and derive relevant formula for this class. Next, in Section \ref{sec:power}
we prove some counter-intuitive properties of the $k$-extendible maps, namely we show that they 
are very powerful in distillation and can distill EPR pairs even from product states.
These peculiarities hold mainly for states which are not of full rank. 
In Section \ref{sec:Werner} we consider distillability of Werner states. 
We provide an analytical formula for fidelity achievable by $1$-extendible maps 
(as well as by the subclass of "measure-and-prepare" maps) from a single copy of 
Werner state. The two classes give the same value of the fidelity. We also present numerical data 
for the fidelity in the case of more copies and $k$-extendible maps with larger $k$.

\section{Formula for Fidelity with $k$-Extendible Maps}
\label{sec:fid}
We start with the following formula for the fidelity achievable for
a given state ρ by a given completely positive, not necessarily
trace-preserving, map Λ:
\begin{align}
  F(ρ,Λ) = \frac{\tr (Λ(ρ) Φ^+)}{\tr (Λ(ρ))}.
\end{align}
where $Φ^+=|ϕ_+⟩⟨ϕ_+|$ and
\begin{align}
  |ϕ_+⟩ = \frac{1}{\sqrt{d}}\sum_{i=0}^{d-1} |ii⟩.
\end{align}
From this formula we obtain

\begin{fact}
\label{fac:fid}
  For any state ρ and any completely positive map Λ, the following
  condition holds
  \begin{align}
    F(ρ,Λ) > α \iff \tr(Λ(ρ) \, M^{(\alpha)}) < 0
    \label{eq:F-gt-cond}
  \end{align}
  where $M^{(\alpha)}=α I - Φ^+$ acts on a two-qubit Hilbert space.

  If additionally ρ is a full rank state then we also have
  \begin{align}
    \label{eq:F-le-cond}
    F(ρ,Λ) < α &\iff \tr(Λ(ρ) \, M^{(\alpha)}) > 0 \\
    \label{eq:F-eq-cond}
    F(ρ,Λ) = α &\iff \tr(Λ(ρ) \, M^{(\alpha)}) = 0.
  \end{align}
\end{fact}

\begin{remark}
  The fidelity is here achievable with some nonzero probability, but
  the probability can be very small, and may depend on α. E.g., when α
  tends to 1, the probability may tend to 0.
\end{remark}

\begin{remark}
  In the singular case when $\tr(Λ(ρ))=0$ the trace expression in
  \eqref{eq:F-gt-cond} is zero and the fact still works.  
  The singular case never happens for the full rank states (if $\Lambda$ does not map all operators to zero)
  \footnote{To see it, it is enough to consider extremal CP maps $\Lambda(\rho)= A\rho A^\dagger$. 
Now, any full rank state satisfies  $\rho\geq  c I$ where $c>0$. Therefore $\Lambda(\rho)\geq c\Lambda(I)= cAA^\dagger$. The operator $cAA^\dagger$ is positive, hence then also its trace does vanish unless 
it is zero operator.} and that is why we have additional conditions
  \eqref{eq:F-le-cond} and \eqref{eq:F-eq-cond} for full rank states.
\end{remark}

We can rewrite $Λ(ρ)$ in \eqref{eq:F-gt-cond}--\eqref{eq:F-eq-cond} by
means of the Choi-Jamio\l{}kowski (CJ) state of the map Λ. Namely, let us
denote the input systems of Λ by $AB$ and the output systems by $ab$.
Then we define CJ state of Λ as follows
\begin{align}
  σ_{A'B'ab}=(\id  ⊗ Λ) \, Φ^+_{A'A} ⊗ Φ^+_{B'B}
\end{align}
where $A',B'$ are of the same dimensions as $A,B$, and $Φ^+$ is
maximally entangled state. (The state is not-necessarily normalized.)
Now it holds \cite{CiDuKraLe01}
\begin{align}
  Λ(ρ_{AB})=d^2 \tr_{AB}(σ_{ABab} \, ρ_{AB}^T ⊗ I_{ab})
  \label{eq:CJ-map}
\end{align}
Using this, we get that the trace expression in \eqref{eq:F-gt-cond}
is proportional to
\begin{align}
  \tr (ρ_{AB}^T ⊗ M_{ab}^α \, σ_{ABab})
  \label{eq:cj}
\end{align}
where the $AB$ is the system on which the state $ρ$ acts, and $ab$ is
a two-qubit system.

This leads to the following version of the previous fact:

\begin{fact}
  \label{fact:fidelity-CJ}
  For any state ρ and any completely positive map Λ, the following
  condition holds
  \begin{align}
    F(ρ,Λ) > α \iff \tr(ρ_{AB}^T ⊗ M_{ab}^α \, σ_{ABab}) < 0
  \end{align}
  where $σ_{ABab}$ denotes the CJ state of Λ.

  If additionally ρ is a full rank state then we also have
  \begin{align}
    F(ρ,Λ) < α &\iff \tr(ρ_{AB}^T ⊗ M_{ab}^α \, σ_{ABab}) > 0 \\
    F(ρ,Λ) = α &\iff \tr(ρ_{AB}^T ⊗ M_{ab}^α \, σ_{ABab}) = 0.
  \end{align}
\end{fact}

\begin{figure}
  \centering
  \includegraphics{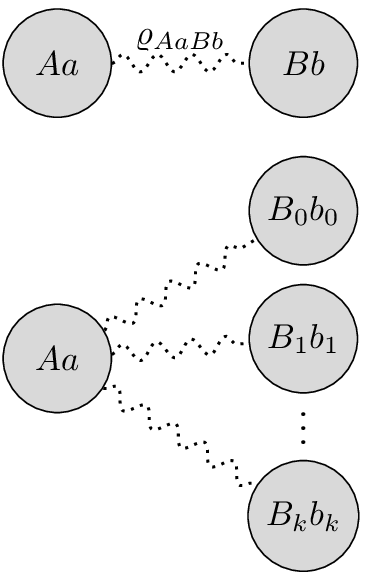}
  \fcaption{$\rho_{AaBb}$ is $k$-extendible state if there exist a state
    $\rho_{AaB_0b_0B_1b_1 … B_kb_k}$ such that $\rho_{AaB_ib_i} = \rho_{AaBb}$
    for $i ∈ \{0, … k \}$.}
    \label{fig:kext}
\end{figure}

We call Λ a $k$-\emph{extendible map} if its CJ state is
a $k$-extendible state. A state $ρ_{AB}$ is $k$-\emph{extendible} (on
Bob's site) if there exist a state $ρ_{AB_0…B_k}$ such that
$ρ_{AB_i} = ρ_{AB}$ for all $i$ from $0$ to $k$.  Analogously we will
say that Λ and $ρ_{AB}$ are $k$-extendible on Alice's site if there
exist a state $ρ_{A_0…A_kB}$ such that $ρ_{A_iB} = ρ_{AB}$ for all $i$
from $0$ to $k$.  We will often consider operators having four
subsystems ($ABab$) instead of two ($AB$) then $A$, $B$, $A_i$ and
$B_i$ will be replaced with $Aa$, $Bb$, $A_ia_i$ and $B_ib_i$ in the
definition of $k$-extendability (see Fig. \ref{fig:kext}). We will use subsystems $Bb$ and
$B_0b_0$ interchangeably and use $Ee$ to denote subsystem
$B_1…B_k,b_1…b_k$, especially when $k=1$.

We use fact \ref{fact:fidelity-CJ} to compute the lower and upper
bounds for the supremum of $F(ρ,Λ)$ over all $k$-extendible maps:

\begin{proposition}
  \label{prop:k-ext}
  For any state $ρ_{AB}$, the supremum of fidelity $F(ρ_{AB}, Λ)$
  achievable by $k$-extendible maps, let us denote it by
  $F_k(ρ_{AB})$, is connected to positivity of some operator, namely
  \begin{align}
    F_k(ρ_{AB}) > α \iff λ_{\min}(Ŝ_k(X^α_{ABab} ⊗ I_{Ee})) < 0
  \end{align}
  and if ρ is a full rank state then also
  \begin{align}
    F_k(ρ_{AB}) < α &\iff λ_{\min}(Ŝ_k(X^α_{ABab} ⊗ I_{Ee})) > 0 \\
    F_k(ρ_{AB}) = α &\iff λ_{\min}(Ŝ_k(X^α_{ABab} ⊗ I_{Ee})) = 0.
  \end{align}
  where $X_{ABab}^α$ is given by
  \begin{align}
    X_{ABab}^α = ρ_{AB}^T ⊗ \M
    \label{eq:X_ABab}
  \end{align}
  subsystem $Ee$ denotes $B_1…B_k,b_1…b_k$ and $Ŝ_k$ denotes the
  symmetrization superoperator
  \begin{align}
    Ŝ_k(X) = \sum_{i=0}^{k} V_{B_0b_0:B_ib_i} X V_{B_0b_0:B_ib_i}
  \end{align}
  where $V_{X:Y}$ swaps subsystems $X$ and $Y$ and for ease of
  indexing we use $B_0$ and $b_0$ to denote $B$ and $b$, respectively.
\end{proposition}

\proof{
  From Fact \ref{fact:fidelity-CJ} we obtain
  \begin{align}
    &F_k(ρ_{AB}) = \sup_{Λ ∈ \{Λ_k\}} F(\Lambda,\rho_{AB}) > α \\
    &\iff \exists_{Λ ∈ \{Λ_k\}} \; F(\Lambda,\rho_{AB}) > α \\
    &\iff \exists_{σ_{ABab} \in \kEXT} \;
    \tr\left[ X_{ABab}^α σ_{ABab} \right] < 0 \\
    &\iff \inf_{σ_{ABab} \in \kEXT}
    \tr\left[ X_{ABab}^α σ_{ABab} \right] < 0.
  \end{align}
  where $\{ Λ_k\}$ denotes the set of all $k$-extendible maps and
  $\kEXT$ is the set of all $k$-extendible states. The right hand side
  can be transformed as follows
  \begin{align}
    \inf_{σ_{ABab} \in \kEXT} \tr\big[ X^α_{ABab} \, σ_{ABab} \big]
    &= \inf_{σ_{ABabEe} ∈ \mathrm{SYM}_k}
    \tr\big[ X^α_{ABab} ⊗ I_{Ee} \, σ_{ABabEe} \big] \\
    &= \inf_{σ_{ABabEe}}
    \tr\big[ X^α_{ABab} ⊗ I_{Ee} \, Ŝ_k(σ_{ABabEe}) \big] \\
    \label{eq:hat-under-trace} &= \inf_{σ_{ABabEe}}
    \tr\big[ Ŝ_k(X^α_{ABab} ⊗ I_{Ee}) \, σ_{ABabEe} \big] \\
    &= \inf_{ψ_{ABabEe}}
    ⟨ψ_{ABabEe}| Ŝ_k(X^α_{ABab} ⊗ I_{Ee}) |ψ_{ABabEe}⟩ \\
    &= λ_{\min}(Ŝ_k(X^α_{ABab} ⊗ I_{Ee}))
  \end{align}
  where $\mathrm{SYM}_k$ is the set of all $k$-symmetric states, the
  equality \eqref{eq:hat-under-trace} comes from $\tr (Λ(A) \, B) =
  \tr (A \, Λ^{†}(B))$ for $Λ ∈ CP$ and from $Ŝ_k^†=Ŝ_k$.

  Thus finally
  \begin{align}
    F_k(ρ_{AB}) > α \iff λ_{\min}(Ŝ_k(X^α_{ABab} ⊗ I_{Ee})) < 0.
  \end{align}

  The proof of the additional condition for full rank ρ is analogous}

\begin{corrolary}
  %For any state ρ one can achieve any $F(ρ,Λ) ≤ α$ by some
  %$k$-extendible map if operator $Ŝ_k(ρ_{AB}^T ⊗ \M ⊗
  %I_{B_1\ldots B_k, b_1\ldots b_k})$ is non-positive.
Given $\alpha>0$, for any state $\rho$  and any $F\leq \alpha$ there exists $k$-extendible map $\Lambda$ 
such that $F(\rho,\Lambda)=F$  if 
operator $Ŝ_k(ρ_{AB}^T ⊗ \M ⊗I_{B_1\ldots B_k, b_1\ldots b_k})$ is non-positive.

 If additionally ρ is a full rank state then 
there exists $k$-extendible map $\Lambda$ 
such that $F(\rho,\Lambda)=F$   
if the smallest eigenvalue of $Ŝ_k(ρ_{AB}^T ⊗ \M ⊗
  I_{B_1\ldots B_k, b_1\ldots b_k})$ is equal to 0.  
\end{corrolary}

\proof{
  From proposition \ref{prop:k-ext} some $F > α$ is achievable by some
  $k$-extendible map $Λ_k$, but then one can use a class of
  $k$-extendible maps $Λ_k^{(p)}$ which with probability $p$ works as
  $Λ_k$ and with probability $1-p$ return a state orthogonal to $Φ^+$
  to obtain any fidelity $F ≤ α$}

Finally, there is the following general question: Can it be, that
probabilistically one can get $F$ arbitrary close to one, but with
probability one, it is not possible? For LOCC, achieving high $F$
probabilistically, means the same deterministically, by law of large
numbers, and postselection. 
%However we do not know whether
%k-extendible maps can be postselected. 
However most likely, the  complement
to $k$-extendible map would not be $k$-extendible anymore, hence the argument would not
be applicable.

%\marginpar{can be postselected vs always are?\tred{\tt o co chodzilo w tym pytaniu?}}

%\tred{\tt write down counter example - nie moge sobie przypomniec,
%chyba damy sobie spokoj}

In other words, while in LOCC case the distillability by means of
trace-preserving maps is equivalent to distillability by a non-trace
preserving ones, for k-extendible maps we do not know if it is
the case. In this paper we do not require preserving of trace, and we
get that the maps considered are very powerful. There is a possibility that
trace-preserving maps are not as powerful (hence more useful for the
problem of distillability) \footnote{Note that the difference between trace-preserving and non-trace-preserving maps is not merely a normalization issue. The condition of trace preserving imposes constraints on the structure of the map.}. However they are much harder to deal with.

\subsection{"Measure and prepare" $k$-extendible maps}

Here we consider a subclass of $k$-extendible maps, which will in
a sense decouple the state ρ from the operator $M^{(\alpha)}$.

\begin{proposition}[The form of are $k$-extendible  M\&P maps]
  Consider any two states $σ_{AB_0\ldots B_k}$ and $σ_{ab_0\ldots
    b_k}$.  We shall denote the reductions $σ_{AB_i}$ by $σ_i^{in}$
  and the reductions $σ_{ab_i}$ by $σ_i^{out}$. Then the following map
  is $k$-extendible: Alice and Bob apply to the given state $ρ_{AB}$
  a global probabilistic POVM whose elements are the states $σ_0^{in},
  \ldots, σ_k^{in}$.  Then given the outcome $i$ they prepare
  (globally) the state $σ_i^{out}$ from the set of states
  $σ_0^{out},\ldots, σ_k^{out}$.  The CJ state of such a map has the
  form $\tr_{B_1\ldots B_k \, b_1\ldots b_k}Ŝ_k(σ_{AB_0\ldots B_k}^T ⊗
  σ_{ab_0\ldots b_k})$.
  \label{prop:pm}
\end{proposition}

\begin{remark}
  Since our maps are not necessarily trace-preserving, the POVM
  elements need not sum to identity.
\end{remark}

\proof{
  We shall prove the case with $k=1$ for clarity (for higher $k$ the
  proof is identical).  Consider a state of the following
  form $\tr_{Ee} Ŝ_1(σ_{ABE}^T ⊗ σ_{abe})$ where $Ŝ_1$
  symmetrizes systems $Bb$ with $Ee$.  Now using \eqref{eq:CJ-map} we
  obtain that the above state is the CJ state of the following map
  \begin{align}
    Λ(ρ_{AB})=\tr(ρ_{AB} σ_{AB}) σ_{ab} + \tr(ρ_{AB} σ_{AE}) σ_{ae}
  \end{align}
  where $σ_{AE}$ and $σ_{ae}$ act on $AB$ and $ab$ systems,
  respectively.  Thus Λ is 1-extendible and its CJ state has the
  postulated form}

\begin{examples}
  The simplest possible map of this form is when there is only one
  state $σ_{out}$.  This means that Alice and Bob remove the initial
  state, and in its place prepare some $k$-extendible state.  Other
  example is when the output states are $σ_0=Φ^+_{ab}$ and
  $σ_i=\frac14I_{ab_i}$ for $i>0$.
%Another way the class can be parametrized is the following.
%We take two states: $σ_{ABB_1\ldots B_k}$ and $σ_{abb_1\ldots b_k}$.
%It is better to use here $B$ instead of $E$
\end{examples}

Using Fact \ref{fact:fidelity-CJ} we obtain that fidelity α is
achievable by measure-and-prepare maps if and only if the following
quantity is non-positive for some states $\sigma_{AB_1\ldots B_k}$ and $\sigma_{ab_1\ldots b_k}$:
\begin{align}
  \sum_{i=1}^k (α-F_i)\tr (ρ_{AB} σ_{AB_i})
  \label{eq:f1f2alpha}
\end{align}
where $F_i$ are overlaps of $σ_{ab_i}$ with $Φ^+$, i.e., $F_i=\tr (σ_{ab_i}Φ^+)$.

Indeed, following the proof of Proposition \ref{prop:k-ext} we obtain the
following criterion:
\begin{proposition}
  Fidelity $F=α$ is achievable if
  \begin{align}
    \inf_{F_1,F_2} λ_{\min}(Z) < 0
    \label{eq:f1f2Z}
  \end{align}
where
\begin{align}
  Z=(α-F_1) ρ_{AB} ⊗ I_E +(α-F_2) ρ_{AE} ⊗ I_B
\end{align}
and infimum runs over all pairs $F_1,F_2$ consistent with the joint state
$σ_{abe}$.
\end{proposition}

\begin{figure}[t]
  \centering
  \includegraphics[width=6cm]{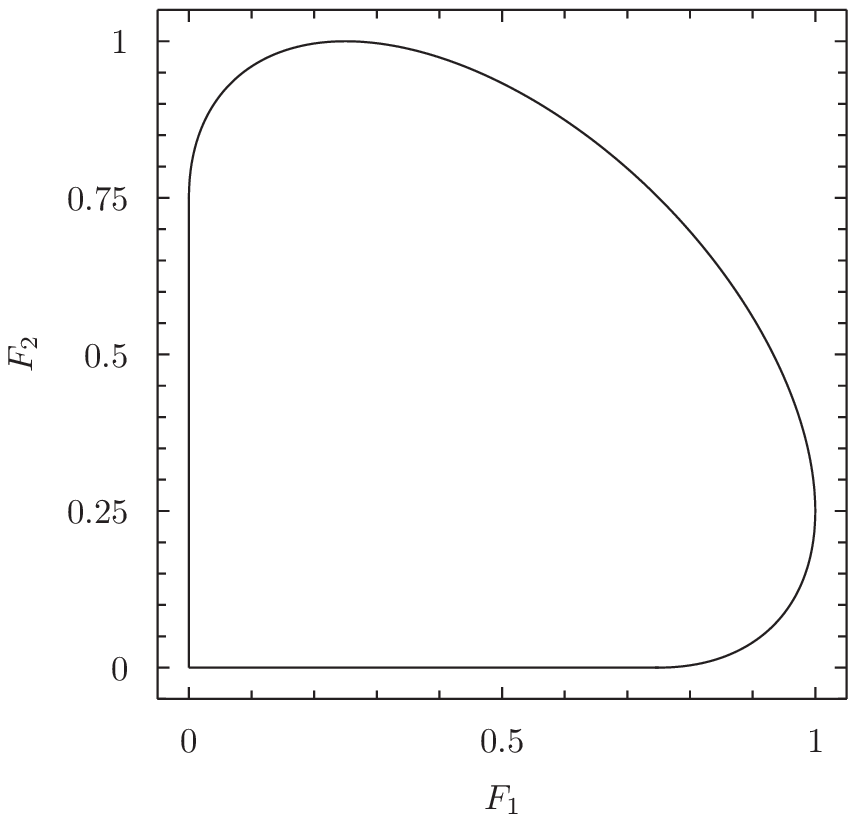}    
  \fcaption{Tradeoff between fidelities $F_1$ and $F_2$ of two-qubit reductions of a tri-qubit
    state. The allowable region is a convex hull of a point 0 (coming from trivial irrep of symmetric group $S_3$)
    and the ellipse described by Eq. \eqref{eq:ellipse}.}
  \label{fig:f1f2}
\end{figure}

\begin{remark}
  It is enough to consider pairs $F_1,F_2$ from the boundary of
  allowed regions.
\end{remark}

For just one extension, the region of possible pairs of fidelities $(F_1,F_2)$ 
\cite{Gisin-rmp-cloning} is a convex hull of a pair $(0,0)$ and the ellipse given by: 
\begin{align}
  y_+^2+\frac13y_-^2 ≤ \frac{1}{16}
  \label{eq:ellipse}
\end{align}
where
\begin{align}
  y_+=(1-F_1-F_2)/2,\quad y_-=(F_1-F_2)/2.
  \label{eq:yplusminus}
\end{align}
This result can also be easily obtained by means of irreducible
representations of symmetric group. To this end it is more convenient
to consider singlet instead of $Φ^+$ and then it is enough to restrict
to the states that are $UUU$ invariant. The allowed region is
depicted on Fig. \ref{fig:f1f2}.
%Clearly to minimize the expression \eqref{eq:f1f2alpha} we need to take pairs fron the
%boundary. This can be found by considering instead overlap with singlet, $UUU$ twirling the state $σ_{ab_1b_2}$
%and using irreps.

%and infimum runs over all $\psi_{ABE}$ and all pairs $F_1,F_2$ lying on
%ellipse given by \eqref{eq:ellipse} (clearly pairs lying inside are subotpimal).

\section{The Power of $k$-Extendible Maps}
\label{sec:power}

The class of $k$-extendible maps converges to the class of
separable operations for large $k$. So one could expect that the
class will not have much more power to distill states than separable
maps, especially when $k$ is large enough. Quite surprisingly, we shall show in
this section that the class is unexpectedly powerful. In particular, in the single copy
case, it can distill perfect EPR pairs from pure product states. 
Moreover for any fixed $k$, with the number of copies $n$ of the
initial state growing to infinity, one can obtain fidelity arbitrarily
close to $1$ from any state different from identity.

\subsection{Single copy: Distillation from product states
and from identity}
\label{subsec:product}

\paragraph*{Distillation from product of a pure and  a mixed state}
Suppose now, that $\varrho^{AB}= \varrho_A ⊗ \varrho_B$ where
$\varrho_B=|0\>\<0|$ and $|0\>$ is an arbitrary fixed vector in system
$B$ (for convenience we assume that it belongs to the basis in which
transpose is taken in formula \eqref{eq:X_ABab}). Then, it is enough
to consider positivity of the operator $X'$
\begin{align}
  X' = Ŝ_k \left(|0⟩⟨0|_{B} ⊗ M_{ab} ⊗ \I_{B_1…B_k, b_1…b_k} \right)
\end{align}
where $M=α \I - Φ^+.$

We shall prove the result for $k=1$. The proof for larger $k$ is
analogous. One finds that
\begin{multline}
  X'= P_B ⊗ P_E ⊗ (M_{ab} ⊗ \I_e+M_{ae} ⊗ \I_b)
  + P_B ⊗ P^\perp_E ⊗ M_{ab} ⊗ \I_e
  + P_B^\perp ⊗ P_E ⊗ M_{ae} ⊗ \I_b
\end{multline}
where $P=|0\>\<0|,P^\perp=\I-P$.
We see that this operator has block diagonal form and the last
block has negative eigenvalue for any $α<1$.  Thus fidelity arbitrary
close to $1$ can be achieved by one-extendible maps. The argument
holds, if $ρ_B$ is proportional to any projector different than
identity.

\paragraph*{Distillation from maximally mixed state.} From the above
consideration, it follows that if $ρ_{AB}=\frac{1}{d_A} I_A ⊗
\frac{1}{d_B} I_B$, then a 1-extendible map can distill it up to
fidelity $F=α$ provided $M_{ab} ⊗ I_e+M_{ae} ⊗ I_b$ is
non-positive. One finds that eigenvalues of this operator are equal to
$\{ 2α,(4α -3)/2, (4α-1)/2\}$. Thus the operator is non-positive, for
$α <3/4$.  Since the state is of full rank, then $F=3/4$ can be
obtained.

For $k$-extendible maps, we need non-positivity of the following
operator $Ŝ_k (M_{ab_0} ⊗ I_{b_1} ⊗ \ldots ⊗ I_{b_k})$, where $Ŝ_k$
symmetrizes over $b_i$'s.  Before we discuss the case of
$k$-extendible maps for $k$ larger than 1, let us describe what
happened here from another perspective.  Namely, the following is
a legitimate $k$-extendible map: to remove the original state, and
bring in a $k$-extendible state $σ_{ab}$. Indeed, the CJ state of such
an operation is given by $σ_{ABab}=\frac{1}{d_A}I_A ⊗ \frac{1}{d_B}I_B ⊗ σ_{ab}$.  (This
is clearly a special case of the measure-and-prepare maps.)  Thus
the fidelity that obviously can be achieved by $k$-extendible maps is
the maximal overlap with $Φ^+$ possible for a $k$-extendible bipartite
state $σ_{ab}$.  However this is related to universal cloning: such
a state would allow to clone with average fidelity (just by
teleporting the state through $k$-extensions of $σ_{ab}$). The problem
of optimal fidelity of universal cloning has been solved e.g., in
\cite{Werner98-cloning}.  Exploiting the formula for "black cow
factor" from this paper, we obtain that the maximal fidelity of
$k$-extendible state on two-qubit system amounts to
\begin{align}
  F_{max}(k)= \frac12 \frac{k+2}{k+1}
  \label{eq:f-max-id}
\end{align}
for $k=1$ we obtain $F=3/4$ which is compatible with the above.  Thus
the maximal fidelity which can be obtained from maximally mixed state
by $k$-extendible operations is given by the formula
\eqref{eq:f-max-id}. Note that for $k=0$ we have perfect fidelity, which is compatible with the fact that if we do not require extendibility at all, we have all quantum operations, including the map that simply brings in a maximally entangled state.

Let us remark, that in the case of separable or LOCC maps, it is
enough that fidelity greater than $1/2$ is obtained to know that
fidelity arbitrarily close to 1 can also be achieved, if we have many
copies of the state. This is because, for states with fidelity grater
than half, there is known distillation protocol, which does the job.
However, this means that to get high fidelity we concatenate two
operations: first we use the one achieving $F>1/2$ and then the
mentioned protocol. However, the set of $k$-extendible maps is not
closed under composition.  This explains why it is possible to obtain
from maximally mixed state fidelity larger than $1/2$.

The examples of product state and maximally mixed state show that the
$k$-extendible maps are not stable with respect to local embedding into
larger Hilbert space. Indeed, the first example goes through, if we
replace $|0\>\<0|$ with whatever projector which does not have full
rank.  Thus a state $\frac1d I ⊗ \frac1d I$, through the second example is not
distillable to maximally entangled state, if it acts on $C^d ⊗
C^d$. However, if we consider the same state on $C^d ⊗ C^{d'}$ where
$d'>d$, fidelity $F=1$ is possible.

\subsection{Single copy: A wide class of states which offer $F=1$}
\label{subsec:F1}
Let us start with a simple condition which, if satisfied, implies that
fidelity $F=1$ can be obtained (with some probability).

\begin{lemma}
\label{lem:strategy1}
%{\bf One-extendible maps}: Given a state $ρ_{AB}$, suppose there exists a state $σ_{ABE}$ such that
%$\tr (ρ_{AB} σ_{AE})=0$ and $\tr (ρ_{AB} σ_{AB})>0$, where $σ_{AB}$,
%$σ_{AE}$ are partial traces of $σ_{ABE}$. Then one can obtain fidelity arbitrary close to $1$ from $ρ_{AB}$
%by one-extendible maps.
Suppose, that there exists a state $σ_{ABB_1\ldots B_k}$ such that
$\tr (ρ_{AB} σ_{AB_i})=0$ and $\tr (ρ_{AB} σ_{AB})>0$, then one can
obtain fidelity $F=1$ from $ρ_{AB}$ by $k$-extendible maps.
\end{lemma}

\proof{
  We shall prove for $k=1$, for larger $k$ proof is similar.  We shall
  use measure-and-prepare strategy (see Prop. \ref{prop:pm}). We take $σ^{in}_1=σ_{AB}$,
  $σ^{in}_2=σ_{AE}$ and $σ^{out}_1=Φ^+_{ab}$, $σ^{out}_2=I_{ab}/4$.
  Then clearly only outcome $i=1$ will be observed, and the output
  state will be $Φ^+$}

\begin{proposition}
  \label{prop:k1F1}
  If a given state $ρ_{AB}$ is not a full rank state then one can
  obtain from a single copy of $ρ_{AB}$ fidelity $F=1$ by means of
  1-extendible maps (either extendible on Bob's or on Alice's site).
  The $F=1$ is achievable by measure-and-prepare maps.
\end{proposition}

\proof{
  We use lemma \ref{lem:strategy1}.  We need to find two bipartite
  states $σ^{(1)}_{AB}$ and $σ^{(2)}_{AE}$, such that they come from some
  joint tripartite state $σ_{ABE}$ and the first of them has nonzero
  overlap with $ρ_{AB}$ and the other one is orthogonal to $ρ_{AB}$.

  If there exists a product state $σ_A ⊗ σ_B$ in the kernel of
  $ρ_{AB}$ then either
  \begin{enumerate}
  \item[(i)] $σ_A ⊗ I_B$ is not in the kernel then we take $σ^{(1)}_{AB}=σ_A ⊗
    \frac{1}{d_B}I_B$ and $σ^{(2)}_{AE}=σ_A ⊗ σ_B$ and by
    lemma~\ref{lem:strategy1} we can achieve fidelity $F=1$ by
    a 1-extendible map extendible on Bob's site; or
  \item[(ii)] $σ_A ⊗ I_B$ is also in the kernel then there must exist $σ'_A$
    such that $σ'_A ⊗ I_B$ is not in the kernel (as $ρ_{AB} \ne 0$)
    and we take $σ^{(1)}_{AB} = σ'_A ⊗ \frac{1}{d_B}I_B$ and $σ^{(2)}_{EB} =
    σ_A ⊗ \frac{1}{d_B}I_B$ and by lemma \ref{lem:strategy1} we can
    achieve fidelity $F=1$ by a 1-extendible map extendible on Alice's
    site.
  \end{enumerate}

  If there is no product state in the kernel then we take any state
  from the kernel as $σ^{(2)}_{AE}$ and $σ^{(1)}_{AB} = σ^{(2)}_A ⊗ \frac{1}{d_B}I_B$
  and by lemma \ref{lem:strategy1} we can achieve fidelity $F=1$ by
  a 1-extendible map extendible on Bobs's site (and also, analogously,
  on Alice's site)}

\begin{proposition}
  \label{prop:kernel}
  If a given state $ρ_{AB}$ has a $k$-extendible state in the kernel
  than one can obtain from a single copy of $ρ_{AB}$ fidelity $F=1$ by
  means of $(k + 1)$-extendible maps (either extendible on Bob's or on
  Alice's site).
\end{proposition}

\proof{
  We extend on the proof of proposition \ref{prop:k1F1}.

  If there is a product state in the kernel of $ρ_{AB}$ then
  proposition \ref{prop:k1F1} gives $σ_{ABE} = σ_A ⊗ σ_B ⊗ σ_E$ which
  by lemma~\ref{lem:strategy1} gives fidelity $F=1$ from a single copy
  of $ρ_{AB}$ by means of 1-extendible maps, either extendible on
  Bob's or on Alice's site.  As $σ_{ABE}$ is a product state one can
  extend it to $σ_A ⊗ σ_B ⊗ σ_E^{⊗ k}$ for any $k$ to obtain by
  lemma~\ref{lem:strategy1} fidelity $F=1$ from a single copy of
  $ρ_{AB}$ by means of $(k+1)$-extendible maps extendible on the same
  site.

  If there is no product state in the kernel any state from the kernel
  can be used as $σ_{AE}$ in the proof of proposition \ref{prop:k1F1}
  so we take the $k$-extendible one which exists by assumption (we
  assume it is extendible on Bob's site for states extendible on
  Alice's site the proof is analogous).  Now, since $σ_{AE}$ is
  $k$-extendible on Bob's site there exists a state $σ_{AB_1…B_{k+1}}
  ⊗ σ_B$ such that $σ_{AB_i} = σ_{AE}$ for $i$ from 1 to $k+1$ and
  thus, (analogously to the proof of proposition \ref{prop:k1F1}) by
  lemma~\ref{lem:strategy1} we can obtain fidelity $F=1$ from a single
  copy of $ρ_{AB}$ by means of $(k+1)$-extendible maps extendible on
  Bob's site}

The above proposition implies the following:
\begin{corrolary}
Any state which has a product vector in kernel, 
can give arbitrary $F<1$ by $k$-extendible (in one of sites Alice or Bob's) maps for all $k$.
\end{corrolary}

\begin{examples}
  Consider projectors onto antisymmetric and symmetric subspaces of $C^d ⊗ C^d$
  given by
  \begin{align}
  P_s=\frac12(I+V),\quad P_{as}=\frac12(I-V),
  \end{align}
  where $V$ swaps the subsystems.  For $P_{as}$ we obtain $P_{as}$,
  that for all dimensions $F=1$ can be obtained for all $k$ (since
  $P_{as}$ has a product state in the kernel),
  irrespectively on what site our maps are extendible (since $P_{as}$
  is symmetric with respect to $A\leftrightarrow B$ exchange).

  In turn, the symmetric projector $P_s$ can give $F=1$ for $k ≤
  d-1$. This is because, its complement, the antisymmetric projector
  is $d-2$ symmetrically extendible for $d≥2$.  But by using
  proposition~\ref{prop:k-ext} for $d=3$ we obtain numerically $F=1$
  for each $k ≤ 4$ and only for $k ≥ 5$ fidelity is decreasing with
  $k$ (figure \ref{fig:c1}). This means that for $k=3$ and $k=4$
  measure-and-prepare maps may be to weak to obtain $F=1$ but general
  $k$-extendible maps still can do this.
\end{examples}

\subsection{Many copies: $k$-extendible maps can distill arbitrary
  state apart from maximally mixed one}
\label{subsec:trivial}

Here we shall show, that the class of $k$-extendible maps can distill
any state apart from maximally mixed one. We shall explain this in the
case of $k=1$. The argument for larger $k$ is analogous.

To this end, we consider
\begin{align}
  X = &\; ρ_{AB}^{⊗ n} ⊗ I_E^{⊗ n} ⊗ M_{ab}^α ⊗ I_e
  + ρ_{AE}^{⊗ n} ⊗ I_B^{⊗ n} ⊗ M_{ae}^α ⊗ I_b
\end{align}
By \mbox{Prop.} \ref{prop:k-ext} arbitrary fidelity $F<α$ can be
obtained if this operator is non-positive for this α. We shall now
argue, that for any $α<1$, there exists $n$ such that this operator is
indeed non-positive. Namely, note that the operator $M$ is
non-positive for such α, hence both $M_{ab} ⊗ I_e$ and $M_{ae} ⊗ I_b$
are non-positive. Furthermore, after normalization, the operators
$ρ_{AB}^{⊗ n} ⊗ I_E^{⊗ n}$ and $ρ_{AE}^{⊗ n} ⊗ I_B^{⊗ n}$ are tensor
powers of two distinct states. Therefore they become more and more
orthogonal for growing $n$. In other words, for $n$ large enough,
there exist orthogonal projectors $P$ and $Q$ which distinguish the
two states with arbitrarily large probability of success.  Thus the
value $\tr (X P_{ABE} ⊗ Φ^+_{ab} ⊗ I_e)$ will be
negative. The exact estimates for the number $n$ of copies needed to
obtain negativity for a fixed α can be obtained from Helstrom
condition for distinguishing two states (i.e., by estimating trace
norm distance between the considered states).  For $k>1$, the same
argument applies: we have $k+1$ different states which are for large
$n$ distinguishable by tomography.

\section{Werner States}
\label{sec:Werner}

In this section we consider $d ⊗ d$ Werner states \cite{Werner1989} in the following
parametrization
\begin{align}
  ρ_W(γ) \sim I + γV
  \label{eq:Werner-IV}
\end{align}
(note that in our formulation of the problem, normalization is not relevant).
The state is separable and PPT if and only if $\gamma\geq -\frac1d$.
In \cite{DurCLB1999-npt-bound} it was shown that for 
$\gamma> \frac12$, obtain fidelity greater than $\frac12$
by just projecting Alice and Bob systems into projectors of rank 2.
Also, conversely, for $\gamma\leq \frac12$, from a single copy, one cannot get 
fidelity greater than $\frac12$. It was then conjectured 
and numerically confirmed that for more copies, we have the same threshold, 
which,in turn, gives rise to conjecture, that Werner states in this region 
are bound entangled. The conjecture is still open. Even for two copies 
there is no analytical proof.

Now, we will compute the maximum 
fidelity achievable by applying a 1-extendible map, i.e., $F(ρ_W(γ),
Λ_1)$. We shall also compute maximum fidelity achievable by means 
of measure-and-prepare 1-extendible maps. It turns out that the two fidelities 
are the same. Finally, we shall present some numerical results for more copies 
and $k$-extendible maps with $k>1$. Note that Werner states, apart from 
the two boundary ones  - the symmetric ($\gamma=1$) and antisymmetric ($\gamma=-1$) 
are full rank. Therefore, due to Fact \ref{fac:fid},
we shall determine achievable fidelity for those states.

\subsection{Analytical solution for distillation of Werner states with 1-extendible maps}

Here we will prove the following proposition
\begin{proposition}
The fidelity achievable by $1$-extendible maps from single copy of the 
Werner state is given by 
\be
F(\rho_W(\gamma),\Lambda_1)=\frac12 + \frac12\sqrt{\frac{1+2γ^2}{4-γ^2}}.
\label{eq:prop1}
\ee
\label{prop:1}
\end{proposition}

\proof{
We will use irreducible representation of symmetric group. To this end, instead of operator $X(α)=Ŝ_k(X^α_{ABab} ⊗ I_{Ee})$ considered in proposition \ref{prop:k-ext} we 
will use similar operator $X'(α)$ where $Ψ^-$ is used instead of $Φ^+$.

Our task is to find such $α_1$ that for any $α < α_1$ the operator
\begin{align}
  X'(α) = X_1 ⊗ Y_1 + X_2 ⊗ Y_2
\end{align}
is not positive. Then $F(ρ_W(γ), Λ_1)=α_1$.

We shall use the following notation
\begin{align}
  \label{eq:Xi-Yi}
  X_1 &= ρ_{AB} ⊗ \I_E, & X_2 &= ρ_{AE} ⊗ \I_B \nonumber \\
  Y_1 &= \Mt_{ab} ⊗ \I_e, & Y_2 & =\Mt_{ae} ⊗ \I_b
\end{align}
where $ρ_{AB}$ and $ρ_{AE}$ are Werner states \eqref{eq:Werner-IV} on
the subsystems given in their subscript.  Here, instead of $M$ we have
put $\Mt=α\I- Ψ^-$, thus all operators given by
\eqref{eq:Xi-Yi} are invariant with respect to unitary operations of
the form $U ⊗ U ⊗ U$.

Clearly, $X_i$ are positive. From Section \ref{subsec:product}, we
also know that for $α ≥ 3/4$, $Y_1+Y_2$ is positive too. But $α=3/4$
can be obtained from any state by $1$-extendible maps (by replacing it
with a suitable symmetrically extendible state, as discussed in
sec. \ref{subsec:product}), so it is enough to work with $Y_1+Y_2$
positive.  Let us remind that all the four operators are invariant
with respect to unitary operations of the form $U ⊗ U ⊗ U$.  Thus,
according to \cite{UUU}, each of them is a linear combination of the
following operators
\begin{align}
  R_+&=\frac16 (\I + V_{(12)}+V_{(13)}+V_{(23)} + V_{(123)}+V_{(321)}),\nonumber\\
  R_-&=\frac16 (\I -V_{(12)}-V_{(13)}-V_{(23)} + V_{(123)}+V_{(321)}),\nonumber\\
  R_0&=\I-R_+-R_-,\nonumber\\
  R_1&=\frac13(2V_{(23)}-V_{(13)}- V_{(12)}),\nonumber\\
  R_2&=\frac{1}{\sqrt{3}}(V_{(12)}-V_{(13)}), \nonumber\\
  R_3&=\frac{i}{\sqrt{3}}(V_{(123)}-V_{(321)}).
\end{align}
Here, $V_{(σ)}$ are swaps, permuting systems according to permutation
σ (written down in terms of cycles).  The operator $R_{\pm},R_0$ are
orthogonal projectors, $R_+,R_-$ being totally symmetric and
antisymmetric ones, respectively. The operators $R_i$, $i=,1,2,3$ have
support on $R_0$. This subspace can be decomposed into tensor product
of two Hilbert spaces, one of them being a qubit. There is
a decomposition such that we have $R_i=I ⊗ σ_i$, where $σ_i$ are Pauli
matrices, $R_0=I ⊗ I_2$.

% fig

So we can write
\begin{align}
  \label{eq:XY}
  X_1&=\sum_i s_i R_i, & X_2&=\sum_i \tilde s_i R_i\nonumber\\
  Y_1&=\sum_i t_i R_i, & Y_2&=\sum_i \tilde t_i R_i
\end{align}
where $i$ runs over $\{0,\ldots, 3, +,-\}$.  Now, since $X_1$ and
$X_2$ are permutations of systems, then $s_\pm=\tilde s_\pm$ and
similarly $t_\pm=\tilde t_\pm$.  Therefore, due to positivity of $X_i$
and $Y_1+Y_2$ we obtain that $s_\pm,\tilde s_\pm ≥ 0$ and $t_+,\tilde
t_+ ≥ 0$.  Moreover $t_-=\tilde t_-=0$, as $Y_i$ act on three qubits,
where the antisymmetric projector is missing.

This implies that the operator $X_1 ⊗ Y_1 + X_2 ⊗ Y_2$ is positive if
and only if the following two qubit operator is positive
\begin{align}
  \frac12(X^q_1 ⊗ Y^q_1+X^q_2 ⊗ Y^q_2)
  \label{eq:2q}
\end{align}
where
\begin{align}
  X^q_1&= \sum_{i=0}^3 s_i σ_i, & X^q_2&= \sum_{i=0}^3 \tilde s_i σ_i\nonumber\\
  Y^q_1&= \sum_{i=0}^3 t_i σ_i, & Y^q_2&= \sum_{i=0}^3 \tilde t_i σ_i
\end{align}
Here $σ_0$ is the identity on the qubit space.  The coefficients $s_i$
etc\mbox{.} can be easily computed,
e.g., $s_i=\tr (X_1 R_i)/\tr (R_i^{†} R_i)$: as each of $X_i$ and
$Y_i$ is a linear combination of the identity and one of $V_{(12)}$
or $V_{(13)}$ so one can first compute $\tr (V_{(12)} R_i)/\tr
(R_i^{†} R_i)$, $\tr (V_{(13)} R_i)/\tr (R_i^{†} R_i)$ and $\tr
(R_i)/\tr (R_i^{†} R_i)$, and compute $s_i$ etc\mbox{.} as the proper
combination of those.

We obtain
\begin{align}
  s_0 &= 1 &
  t_0 &= -\frac12 + α \\
  s_1 &= \frac12\gamma &
  t_1 &= -\frac14 \\
  s_2 &= -\frac{\sqrt3}{2}\gamma &
  t_2 &= \frac{\sqrt3}{4} \\
  s_3 &= 0 &
  t_3 &= 0 \\
  \tilde{s}_i &= \begin{cases}
    s_i & i ∈ \{0, 1, 3\} \\
    -s_i & i=2
  \end{cases} &
  \tilde{t}_i &= \begin{cases}
    t_i & i ∈ \{0, 1, 3\} \\
    -t_i & i=2
  \end{cases}
\end{align}

The two qubit operator \eqref{eq:2q} has in terms of the coefficients
$s_i$ and $t_i$ the following form
\begin{multline}
  \label{eq:st-matrix}
  \frac12(X_1^q ⊗ Y_1^q + X_2^q ⊗ Y_2^q) =
  \begin{bmatrix}
    s_0 t_0 & s_0 t_1 & s_1 t_0 & s_1 t_1 - s_2 t_2 \\
    s_0 t_1 & s_0 t_0 & s_2 t_2 + s_1 t_1 & s_1 t_0 \\
    s_1 t_0 & s_2 t_2 + s_1 t_1 & s_0 t_0 & s_0 t_1 \\
    s_1 t_1 - s_2 t_2 & s_1 t_0 & s_0 t_1 & s_0 t_0
  \end{bmatrix}
\end{multline}
and has the following eigenvalues
\begin{align}
  λ_{1,2} &=
  \pm\sqrt{s_2^2 t_2^2 + (s_0 t_1 - s_1 t_0)^2} - s_1 t_1 + s_0 t_0 \\
  λ_{3,4} &=
  \pm\sqrt{s_2^2 t_2^2 + (s_0 t_1 + s_1 t_0)^2} + s_1 t_1 + s_0 t_0.
\end{align}
We have to find $α_1$ such that for any α less then $α_1$ at least one
of eigenvalues $λ_2 ≤ λ_1$ and $λ_4 ≤ λ_3$ is negative. It turns out
that both $λ_2$ and $λ_4$ are zeroed for the same $α=α_1$ which is the
greater of the roots of the equation
\begin{align}
  s_2^2 t_2^2 + (s_0 t_1 \mp s_1  t_0)^2 = (s_1 t_1 \mp s_0 t_0 )^2
\end{align}
which is (up to the normalization) equivalent to a quadratic equation
\begin{align}
  \label{eq:quadratic}
  (16 - 4 γ^2) α^2 - (16 - 4 γ^2) α + (3 - 3 γ^2) = 0.
\end{align}
The greater of the solutions of \eqref{eq:quadratic} has the form
\begin{align}
  \label{eq:alpha1}
  α_{\max} &= \frac12 + \frac12\sqrt{\frac{1+2γ^2}{4-γ^2}}.
\end{align}
Thus, using Fact \ref{fac:fid}, we obtain for all Werner states 
excluding the two boundary ones, the achievable fidelity is  $F(ρ_W(γ), Λ_1)=α_{\max}$ where $α_{\max}$ is given by \eqref{eq:alpha1}. The boundary states 
(those with $\gamma=\pm1$) are not of full rank, hence 
the very fact that $\alpha_{\max}=1$ implies only that
fidelity arbitrary close to $1$ can be obtained.
However, from Proposition \ref{prop:k1F1} we know that 
in the case of $1$-extendible maps, for any state which is not of full rank, 
fidelity $1$ can be achieved}

Due to using of the $I + γV$ parametrization of the Werner state the
solution \eqref{eq:alpha1} has a simple dimension independent form and
is a symmetric function.

One can transform \eqref{eq:alpha1} to the following parametrization of
the Werner state
\begin{align}
  ρ_W(p) = p \frac{P_s}{d_s} +(1-p) \frac{P_{as}}{d_{as}}
  \label{eq:param}
\end{align}
where $P_s$ and $P_{as}$ are, respectively, projectors on to the
symmetric and antisymmetric subspaces, and $d_s$ and $d_{as}$ are their
dimensions.  The transformation can be done using the substitution
\begin{align}
  γ = -\frac{2dp-d-1}{2p-d-1}.
\end{align}
In particular for $d=4$ we obtain
\begin{align}
  α_{\max} = \frac12 + \sqrt{\frac14 - \frac{15p(1 - p)}{25 - 16p^2}}.
  \label{eq:F-e1c1}
\end{align}

\subsection{Distillation of Werner states by 1-extendible measure-and-prepare maps}

We now consider a single copy of Werner state and the "measure and prepare" 1-extendible maps. We shall show that the  fidelity is the same as in the case of all $1$-extendible maps. 
To this end, we need to find
minimum eigenvalue of the operator $Z$ given by
\eqref{eq:f1f2Z}. Using irreducible representations of symmetric
group, we can write $Z$ as
\begin{align}
  Z=(α-F_1)X_1 +(α-F_2)X_2
\end{align}
where $X_i$ are given by \eqref{eq:XY}.
We obtain that
\begin{align}
  Z_q=\sum_{i=0}^3 \beta_i σ_i
\end{align}
where $\beta_i$ are given by
\begin{align}
  \beta_i=(α-F_1)s_i +(α-F_2)\tilde s_i
\end{align}
Recall, that $s_3=\tilde s_3=0$.
Here $Z_q$ denotes the restriction of $Z$ to the qubit, similarly as it was for for $X^q_i$ and $Y^q_i$.

The operator is positive if and only if
\begin{align}
  \sum_{i=1}^3 \beta_i^2 ≤ \beta_0^2
\end{align}
We have now to check this inequality for possible pairs of fidelities. 
However, it is enough to restrict to extremal points, and the pair $(0,0)$ need not be taken into account,
so that we need to take pairs that belong to the ellipse \eqref{eq:ellipse}.
Then, if we put equality in the above formula, there are the following two solutions:
%\begin{widetext}
%\begin{align}
%α_{1,2}=\frac{1}{2} \left(-2 y_++1\pm\frac{2 \sqrt{3} y_-^2 (-2 d
%   p+d+1)^2}{\sqrt{y_-^2 (-(2 (d-2) p+d+1)) (-2 d p+d+1)^2 (2
%   (d+2) p-3 (d+1))}}\right)
%\end{align}
%\end{widetext}
%which can be written as follows
\begin{align}
  α_{1,2}=\frac12-y_+\pm|y_-| f(p,d)
\end{align}
where
\begin{align}
  f(p,d)= \frac{\sqrt{3}|-2 d p+d+1|}{\sqrt{((2 (d-2) p+d+1))(3(d+1)-2(d+2)p)}}
\end{align}
and
\begin{align}
  y_+^2+\frac13 y_-^2=\frac{1}{16}
\end{align}
with $y_\pm$ given by \eqref{eq:yplusminus}.
We have now to maximize the α's over $y_+,y_-$ satisfying the above constraints. This gives
\begin{align}
  α_{max}=\frac{1}{4} \left(\sqrt{3 f(p,d)^2+1}+2\right)
\end{align}
which, once applied \eqref{eq:param}, is exactly the same as the fidelity achievable with
the general $1$-extendible map given in Prop. \ref{prop:1}.

\paragraph*{Two copies}

Note, that in $I+γV$ parametrization, $α_{max}$ does not depend on
dimension, which is partially responsible for its very simple form. However 
the parametrization does not help much for two
copies -- we are able to obtain the expression for eigenvalues of the
expression for two copies
\begin{align}
  X_1^q ⊗ X_1^q ⊗ Y_1^q+X_2^q ⊗ X_2^q ⊗ Y_2^q
\label{eq:2copies}
\end{align}
in terms of $s_i$ and $t_i$ but these are huge expressions and even
after substituting $s_i$ and $t_i$, i.e., in terms of α and γ they
stay huge. 

\begin{figure}
  \centering
  \includegraphics{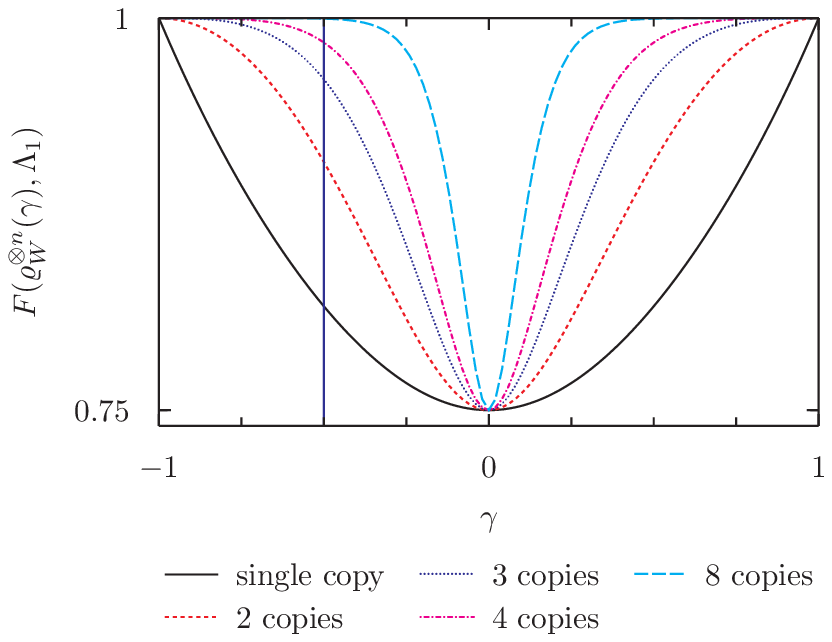}
  \fcaption{Fidelity achievable by 1-extendible maps $\Lambda_1$ on $n$ copies of
    Werner state for $n=1,2,3,4,8$. $I + γV$ parametrization is used. One can observe that
    given sufficiently many copies all states except maximally mixed
    one are distillable with 1-extendible map with arbitrary
    fidelity. The plots are done for arbitrary $d$.}
    \label{fig:1ex_n_copies}
\end{figure}

\begin{figure}
  \centering
  \includegraphics{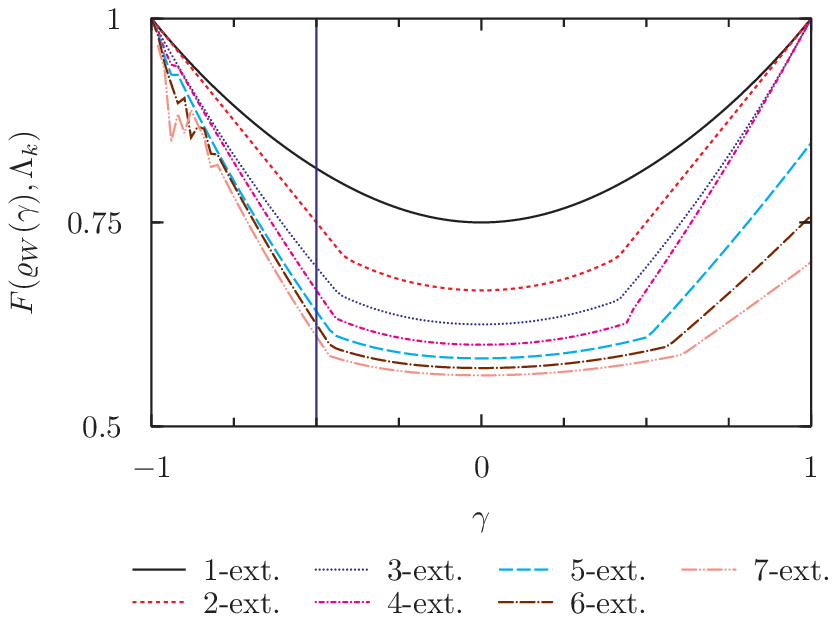}
  \fcaption{Fidelity achievable by means of $k$-extendible maps $\Lambda_k$ ($k=1,\ldots,7$) 
  for a single copy of Werner state with $d=3$.}
  \label{fig:c1}
\end{figure}

\begin{figure}
  \centering
  \includegraphics{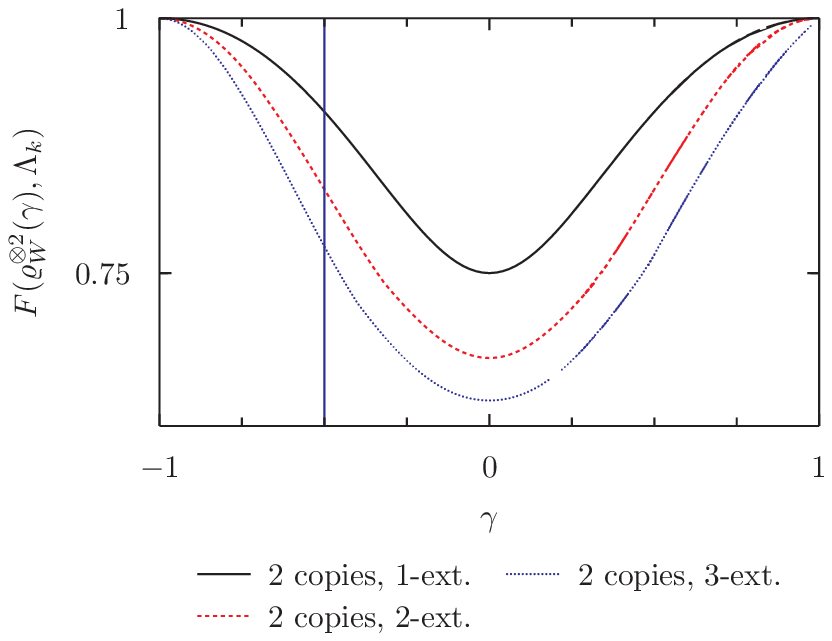}
  \fcaption{Fidelity for two copies of Werner state with $d=3$, and $k=1,2,3$ extendible maps $\Lambda_k$. 
  The larger the number $k$, the lower the curve.}
    \label{fig:many-ext}
\end{figure}

\begin{figure}
  \centering
  \includegraphics{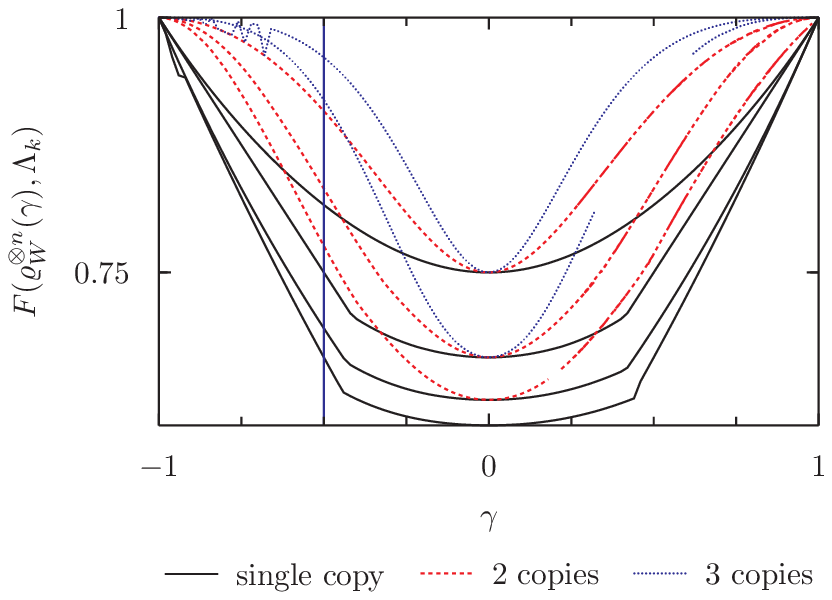}
  \fcaption{Fidelity achievable by action of $k$-extendible maps on $n$ copies of Werner states with $d=3$,
  for $k=1,2,3,4$ and $n=1,2,3$ (for given number of copies, the larger $k$ the lower is the plot). 
  For maximally mixed state ($\gamma=0$) the number of copies does not matter, and the fidelity 
  is given by $F=\frac12\frac{k+2}{k+1}$ (See Eq. \ref{eq:f-max-id}).}
  \label{fig:all}
\end{figure}

\subsection{More copies and more extensions}
We have obtained numerical results for larger number of copies and 
$k$-extendible maps with larger $k$. We present the results on 
subsequent figures. 
%On all figures, the vertical lines are the same:
%the right one separates entangled and NPT states to the left 
%from separable and PPT states to the right. 
%The left line denotes the conjecture boundary between distillable (to the right)
%and non-distillable states (to the left). 
On all figures, the vertical line $\gamma=-1/2$ denotes 
the conjectured boundary between distillable (to the left)
and non-distillable states (to the right). The boundary between separable (to the right) 
and entangled states (to the left) is for $\gamma=0$. 
On Fig. \ref{fig:1ex_n_copies} we present the plot for exemplary 
numbers of copies up to $n=8$, and for $1$-extendible maps. 
For $n=1$ we use the analytical solution 
\eqref{eq:F-e1c1} while for more copies we do numerical computations
i.e. we are diagonalizing the operators of the sort of
\eqref{eq:2copies} with number of $X$'s equal to number of copies.
The plot confirms the result of Sec. 
\ref{subsec:trivial}: for larger and larger number of copies, 
the fidelity of any state but the maximally mixed one tend to $1$.

We have also done exemplary numerical calculations for more extensions and more copies. 
On Fig. \ref{fig:c1} we consider single copy, and $k$-extendible maps up to $k=7$. 
We see that up to $k=4$ the fidelity for symmetric state (one with $\gamma=1$)
has fidelity equal to $1$, and only for $k\geq 5$ the fidelity drops down. 
As discussed in Sec. \ref{subsec:F1}, we have analytical proof that $F=1$ 
for $k\leq 2$, while the cases $k=3,4$ are still not fully understood.
We also can see, that up to $k=4$ the plots are symmetric with respect to 
maximally mixed state ($\gamma=0$). 
This means that for the classes of $k$-extendible maps up to $k=4$,
entanglement/separability property of Werner states is completely irrelevant.

Note also, that once $k$ is growing two cusps are forming: 
the right one will materialize in the coordinates $(\alpha=\frac12, \gamma=1)$ and will mean, 
that all state with $\gamma>0$ are not distillable. 
The left one tends to $(\alpha=\frac12, \gamma=-\frac12$,
where it will constitute the boundary of distillable region according to \cite{DurCLB1999-npt-bound}.
Finally, on Fig. \ref{fig:many-ext} we consider two copies and $k$-extendible maps with $k=1,2,3$
for $d=3$. 
and on Fig. \ref{fig:all} 
we put all the plots together, to visualise, what happens if we change both the number of copies of the state and the number of extensions for the maps. 

On figures \ref{fig:c1} and \ref{fig:all} some curves are ragged for
values around zero.  This happens if computation of eigenvalues of
some of the diagonal blocks of the matrices involved fails. Then the
computed fidelity is underestimated.

%\section{Plots}

%For all of the plots presented below $A=3$ and $a=3$.  Matrices
%$X_{c,k}(p, α)$ have many blocks of
%different dimension, the eigenvalues for the plot are computed
%numerically in each block separately.

%\begin{center}
%\small
%\begin{tabular}{crrr}
%  \hline
%  $c, k$ & 1,5  & 2,3 & 3,1 \\ \hline
%  total dimension
%  & 279,936 & 1,889,568 & 157,464 \\
%  {max block dim. $X_{c,k}$}
%  & 7350  & 9000  & 648  \\
%  {number of blocks $X_{c,k}$}
%  &  288  & 2646  & 4000  \\
%  \hline
%\end{tabular}
%\end{center}

\nonumsection{Acknowledgments}
\noindent
FB is supported by a "Conhecimento Novo" fellowship from the Brazilian agency Funda\c{c}\~ao de Amparo a
Pesquisa do Estado de Minas Gerais (FAPEMIG) and by the National Research Foundation and the Ministry of Education, Singapore. 
\L{}. P. is supported Polish Ministry of Science and Higher
Education grant \mbox{no.} 3582/B/H03/2009/36. M.H. and \L{}. P. are supported 
by the European Commission through the Integrated Project FET/QIPC QESSENCE. 
G.S. is supported by DARPA QUEST contract HR0011-09-C-0047.  F.B. and G. S. thank the 
hospitality of  National Quantum Information Centre of Gda\'nsk where this work 
was initiated.  F.B., M.H., and G. S.  thank the hospitality of Institute Mittag Leffler 
within the program "Quantum Information Science" (2010), where part of this work was done.

\bibliographystyle{hieeetr}
% %\bibliography{rmp14-hugekey}
% %\bibliography{rmp15-hugekey}

\nonumsection{References}
\noindent

%\bibliography{rmp12-hugekey,extendible}

\end{document}